\documentclass[prb,twocolumn,amsmath]{revtex4}
\input {epsf.sty}
\usepackage{epsfig}
\usepackage{amssymb}
\begin{document}
\title{Intrinsic Impurity in the High Temperature Superconductor
Bi$_2$Sr$_2$CaCu$_2$O$_{8+\delta}$}
\author{Bo Chen, Sutirtha Mukhopadhyay, W. P. Halperin, }
\affiliation{Department of Physics and Astronomy,\\
     Northwestern University, Evanston, Illinois 60208}
\author{Prasenjit Guptasarma}
\affiliation{Department of Physics,\\
     University of Wisconsin-Milwaukee, Wisconsin, 53211}
\author{D. G. Hinks}
\affiliation{Materials Science and Technology Division,\\Argonne  National Laboratory,
     Argonne, Illinois, 60439}
\date{Version \today}

\begin{abstract} {The ${}^{17}$O NMR spectra of Bi$_2$Sr$_2$CaCu$_2$O$_{8+\delta}$ (Bi-2212) single
crystals were measured in the temperature range from 4 K to 200 K and magnetic fields from 3 to 29
T, reported here principally at 8 T. The NMR linewidth of the oxygen in the CuO$_{2}$ plane was
found to be magnetically broadened  with the temperature dependence of a Curie law where the Curie
coefficient decreases with increased doping. This inhomogeneous magnetism is an impurity effect
intrinsic to oxygen doping and persists unmodified into the superconducting state.}
\end{abstract}

\maketitle

\vspace{11pt}

Cuprate high temperature superconductivity emerges from the insulating antiferromagnetic parent
compound by chemically doping the system with extra charge. In the case of
Bi$_2$Sr$_2$CaCu$_2$O$_{8+\delta}$ (Bi-2212) this is
 accomplished by inserting $\delta$-oxygen non-stoichiometrically in the crystal structure.  The
doped holes drive the system away from antiferromagnetism giving rise to superconductivity in a
classic dome shaped region in the temperature-dopant phase plane.  Hole doping controls the
balance between magnetism and superconductivity. But scanning tunneling microscopy (STM)
measurements show that doping also produces nano-scale electronic disorder\cite{Pan01,McE05}.  
Neither the doping dependence of T$_{c}$ nor depairing effects from spatial electronic
inhomogeneities are  well-understood as yet.  Nonetheless, it seems from nuclear magnetic resonance
(NMR) measurements that magnetic moments emerge  where there is  local suppression of the
superconducting order parameter\cite{All91}.    

We find  that oxygen doping  also generates  magnetic moments in the copper oxygen plane and that
this local moment behavior persists deep into the superconducting state.  The effect of these impurities is
reduced with increased doping as the system is displaced further from the antiferromagnetic region of the
phase diagram. It remains an open question as to how these moments are formed and if their effect on
superconductivity is deleterious.

%
%

\begin{figure}[h]
\vspace{0.3in}
\includegraphics[width=8 cm]{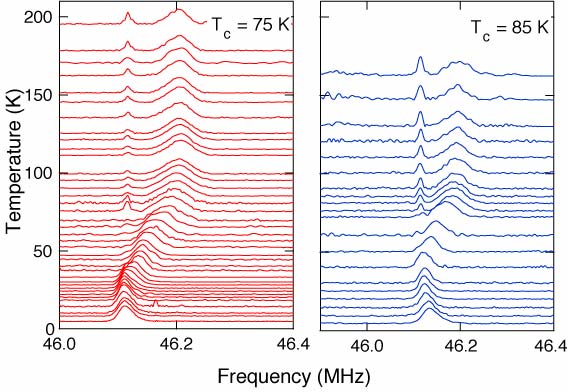}
\suppressfloats [b]
\caption[NMR spectrum]{The central transition of $^{17}$O NMR spectra of two over doped Bi-2212
single crystals. The left plot is for T$_{c}$ = 75 K and the right is for T$_{c}$ = 85 K. The
narrow line is for oxygen in the SrO plane, O(2), and the broader line is for oxygen in the
CuO$_{2}$ plane. The magnetic field is $H$ = 8 T.}
\label{Fig1}
\end{figure}
%

STM and NMR are both microscopic probes of electronic structure and have been used to study the
effect of  chemical impurities in the copper oxygen
plane\cite{Pan01,McE05,All91,Ish93,Mah94,Bob97,Bob99,Jul00,Miy98,Pan00,Hud01,Ver04,Sof06} in cuprates. STM
has powerful capability as a spectroscopic tool but is limited to the charge degrees of freedom within
$\approx 1$ nm of the free surface\cite{Pan01,McE05,Miy98,Pan00,Hud01,Ver04}.  NMR is a complementary
probe\cite{All91,Ish93,Mah94,Bob97,Bob99,Jul00} with spin-sensitivity greater than a penetration
depth from the surface with selectivity for different atoms in the structure,
oxygen in our case.  It is easily studied as a function of temperature and over a wide range of magnetic
fields.  Alloul and Bobroff {\it et al.}\cite{All91,Bob97,Bob99} have used
$^{17}$O to investigate local magnetic field inhomogeneity in the CuO$_{2}$ plane for chemically
doped
$c$-axis aligned powders of YBa$_{2}$Cu$_{3}$O$_{7-\delta}$ (YBCO).  They found a universal
behavior, a Curie temperature dependence, of the
$^{17}$O NMR linewidth for various impurities (Ni, Zn, and Li) that substitute for copper in the
CuO$_{2}$ plane. Alloul {\it et al.} proposed that the chemical substitution for copper atoms in the CuO$_2$
plane  breaks the in-plane, spin-singlet correlations of neighboring copper atoms, giving rise to an
uncompensated local magnetic moment that couples through an oscillating hyperfine interaction to the
$^{17}$O NMR such as might be expected for the RKKY interaction.  As a result the width of the NMR spectrum is
increased symmetrically and proportional to the magnetization of the local moment having the
temperature dependence of a Curie law. Here we report similar behavior to be
intrinsic to oxygen doped Bi-2212.

We have performed  $^{17}$O NMR on three crystals of Bi$_2$Sr$_2$CaCu$_2$O$_{8+\delta}$ (Bi-2212)
with differing amounts of non-stoichiometric $\delta$-oxygen  varying from near-optimal to
over-doped.  The crystals were grown by the floating-zone technique using a two-mirror image furnace.
Cationic homogeneity was ensured by maintaining the same stoichiometry on the top and bottom rods and a
very slow growth rate of 0.1 mm/hr. The crystals were oxygen isotope exchanged in a circulating gas stream
at $\approx 1$ bar, enriched to 70\% $^{17}$O, at 650 $^{\circ}$C for 48 hours.  After exchange
the crystals were annealed; for T$_{c}$ = 75 K (strongly overdoped) we chose 450 $^{\circ}$C 
at 1 bar for 150 hr. Similar procedures \cite{Wat97} at lower partial pressures and higher annealing
temperatures were used to reduce the doping for the other crystals giving transition temperatures
determined from low field susceptibility, T$_{c}$ = 90, 85, and 75 K, and with weights 34 mg, 48 mg, and 28
mg, respectively.   Prior to
$^{17}$O exchange the T$_{c}$ of these crystals was 95 K with optimal doping, the highest yet reported,
indicative of high cationic homogeneity and chemical purity.  After exchange
the concentration of
$^{17}$O was estimated to be
$\approx 60\%$.  As a procedural check we processed two crystals at the same time giving 
T$_{c}$ = 85 K. Then one of these was reannealed to give T$_{c}$ = 90 K.  The NMR spectra of both crystals
were  found to be identical before the final annealing,  providing direct
evidence that the effects we report here are solely attributable to doping.

Fourier transform of the NMR echo from  $\pi/2$ -
$\pi$ spin-echo sequences gave the spectra
of the central transitions $\left\langle {-{1 \over 2}\leftrightarrow +{1 \over 2}} \right\rangle$
shown in Fig. 1. Measurements were performed as a function of
temperature from 4 to 200 K over a range of magnetic field from 3 to 29 T parallel to the
$c$-axis.  Here we discuss our most extensive measurements of Knight shift and
linewidth (Fig. 2 and 3) as a function of temperature and doping in a magnetic field of $H$ = 8
T. 

There are two distinguishable oxygen sites in the NMR spectra identified in previous
work\cite{Tro90} as the oxygen in the CuO$_2$ plane, O(1), and the oxygen in the SrO plane, O(2).
The central transitions are shown in Fig. 1 where the broad resonance is from O(1) and a narrow,
partially saturated, resonance is from O(2). The much narrower NMR spectrum from O(2) indicates a
more homogeneous electronic environment in the SrO plane compared to the conduction plane. 
Additionally, the position of the O(2) resonance does not change significantly with temperature
and its spin-lattice relaxation time is an order of magnitude longer as a consequence of the fact
that O(2) is more weakly coupled to the electronic excitations in the CuO$_{2}$ plane, as compared
to O(1).  In contrast, the O(1) NMR central transition, Fig. 1, is relatively broad.  The central
transition linewidth is similar to its satellites and proportional to magnetic
field\cite{Che07} demonstrating that the spatial inhomogeneity is  from local magnetic fields and that
quadrupolar broadening from a distribution of electric field gradients in the copper oxygen plane is
insignificant.   We exploit the long spin-lattice relaxation time of O(2) to suppress it using fast pulse
repetition for temperatures below 40 K and we explicitly subtract it from the spectrum at higher temperatures
where it can be resolved.

\begin{figure}[h]
\vspace{0.3in}
\includegraphics[width=7 cm]{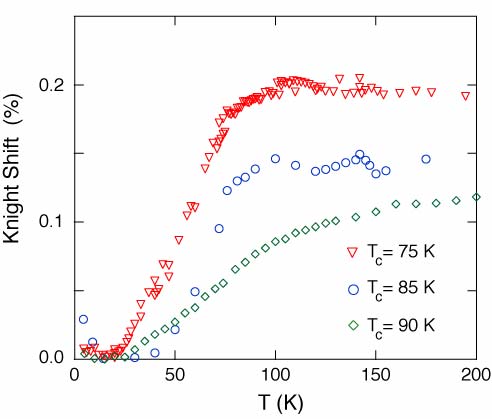}
\suppressfloats [b]
\caption[Temperature dependence of the O(1) Knight shift]{Temperature  dependence of the Knight
shift defined to be the first moment of the
$^{17}$O(1) central transition of each of three crystals.}
\label{Fig2}
\end{figure} 

\begin{figure}[h]
\vspace{0.3in}
\includegraphics[width=8 cm]{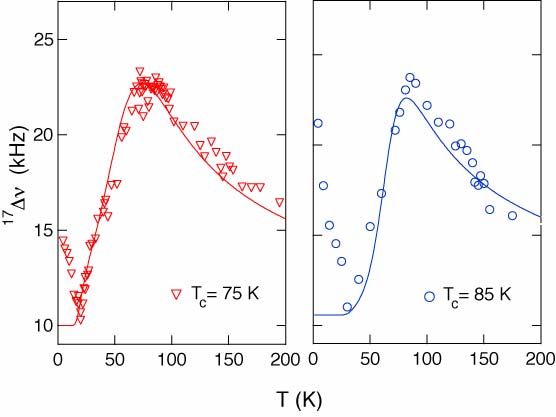}
\suppressfloats [b]
\caption[Temperature dependence of the O(1) linewidth]{The linewidth of the
$^{17}$O(1) central transition, calculated as the square root of the second moment as
a function of temperature.   The solid curves are from Eq. 1. The sharp upturn in data near 30 K
marks the vortex freezing transition\cite{Che07}.}
\label{Fig3}
\end{figure}

In the superconducting state the frequency of the O(1) resonance peak  decreases with temperature
below the superconducting transition, Fig. 2, approaching its zero Knight shift position, $K(0)$, near 30
K. This reduction is expected for a spin-singlet superconductor where $K(T)$ follows the
temperature dependence of the spin susceptibility.  Well above the superconducting transition
temperature, $K(T)$ approaches a  temperature independent value that increases with hole doping as shown
in Fig. 4. This general trend is expected; in a simple metal the spin susceptibility is proportional to
the electronic density of states.

The linewidth of O(1), Fig. 3, has an unusual Curie temperature
dependence in the normal state.  This observation and its extension to the superconducting state
is the main focus of our work.  Such a Curie behavior comes from paramagnetic moments that produce
a static inhomogeneous distribution of magnetic field throughout the sample, thereby broadening
the NMR spectrum increasingly with decreasing temperature as was first observed by Alloul {\it et
al.}\cite{All91} in the normal state of cation-substituted YBCO.  In the
superconductive state, vortex supercurrents generate an inhomogeneous field distribution that
make a quantitative interpretation more complex\cite{Sof06}, except in the vortex liquid region
where rapid vortex motion on the NMR time scale, averages this inhomogeneity to zero. In the case
of Bi-2212, in contrast with YBCO, the vortex liquid state occupies a wide temperature range due
to its high anisotropy\cite{Wat01} permitting us to extend measurement of local moment impurities
well below T$_{c}$. In our samples vortex freezing is identified with the sharp
increase of the linewidth with decreasing temperature near 30 K, Fig. 3,   from which we have
determined\cite{Che07} the freezing phase diagram up to $H \approx$ 30 T.  

The strong decrease in linewidth with decreasing temperature that we observe in the superconducting
state, Fig. 3,  must therefore be a consequence of some combination of local moment behavior and
superconductivity.  The narrowing of the spectrum seems, at first sight, to parallel that of the
temperature dependent Knight shift suggesting that either the conduction electrons are
crucial to the formation of these local moments or, at the very least, our sensitivity to them
through the hyperfine interaction is interrupted as conduction electrons condense into a singlet,
Cooper-pair, state.  In fact, we have found that the relation between the linewidth and Knight shift
over the whole temperature range can be represented by a simple phenomenological expression:

\begin{eqnarray}
\Delta\nu(T)=\Delta\nu_{0}+K(T){\cdot}H{\cdot}C/T ,
\end{eqnarray}
where $\Delta\nu$ is the observed linewidth, $K(T)$ is the Knight shift,
$T$ is the temperature, and $C$ is a Curie constant. $\Delta\nu_{0}$ denotes the temperature
independent intrinsic linewidth determined by extrapolating the decreasing linewidth to a
temperature where $K(T)$ approaches zero. For each sample this is proportional to the field, $\approx$ 1
kHz/T.  Our fits to Eq.1  are given by the solid curves in Fig. 3 with only the Curie constant as an
adjustable parameter.  Here  $K(T)$ in the formula is replaced by a numerical representation of the data
in Fig. 2.  The resulting values for
$C^{-1}$ are plotted in Fig. 4 and appear to vary with hole concentration in a manner similar to the
high temperature value of the Knight shift,
$K(T>>T_{c})$. 

We can also make a direct comparison of our raw data with Eq. 1.  In Fig. 5 we plot, for each
measured O(1) spectrum, the ratio of its first moment to the square root of its second moment (subtracting
the fixed background contribution).  The raw data in this form indicates a Curie law, Eq. 1, if it is a
straight line passing through the origin. We conclude that this is the correct temperature dependence of
the local moment behavior for both normal and superconducting states for the two overdoped crystals. For
the optimally doped crystal we suspect that the pseudogap contributes to the temperature dependent Knight
shift sufficiently that there are deviations from the Curie law behavior.  
\begin{figure}[h]
\vspace{0.3in}
\includegraphics[width=6.5 cm]{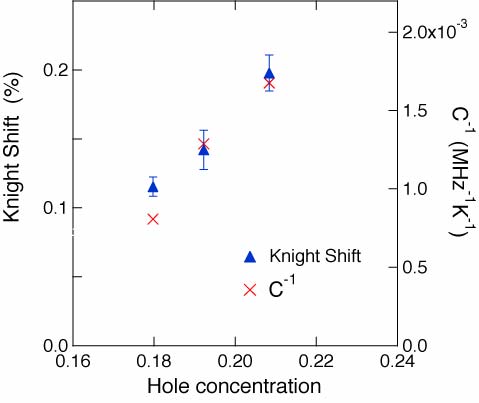}
\suppressfloats [b]
\caption[Knight shift and hole concentration relation]{Knight shift and  inverse Curie constant as
a function of hole concentration.  The hole concentration in the CuO$_2$ plane is related to T$_c$
by\cite{Pre91}: T$_c$/T$_{cmax}$ = 1 - 82.6 (p - 0.16)$^2$. The non-stoichiometric oxygen content,
$\delta$, was determined indpendently from  annealing conditions\cite{Jea00} giving values of
$\delta$ close to the hole concentration. }
\label{Fig4}
\end{figure}
\begin{figure}[h]
\vspace{0.3in}
\includegraphics[width=8 cm]{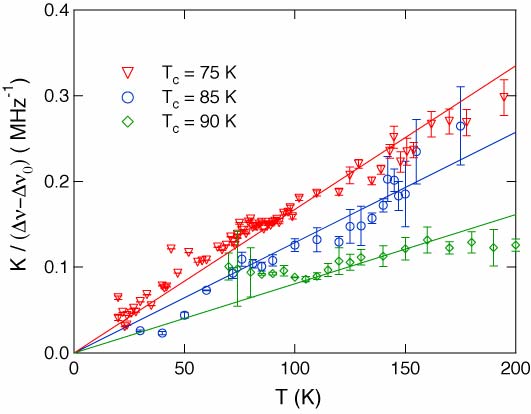}
\suppressfloats [b]
\caption[Temperature dependence of the O(1) linewidth]{The Knight shift (first moment of the
spectrum) divided by the linewidth (defined as the square root of the second moment),
K/($\Delta$$\nu$-$\Delta$$\nu$$_{0})$, is plotted against temperature. A straight line through the origin
indicates a Curie law with slope the inverse Curie constant.}
\label{Fig5}
\end{figure}

The magnetization of YBCO samples with Ni, and Zn substituted for copper in the
CuO$_{2}$-plane have a Curie temperature dependence in the normal
state\cite{Coo91,Men94,Bobthesis} that is correlated with a corresponding temperature dependence of the
NMR linewidth, indicating the existence of paramagnetic moments\cite{All91,Bob97,Bob99,Jul00}. The
similar behavior that we observe
we associate with oxygen doping. However, if chemical impurities were to be present, in order to broaden
the NMR spectra as reported for YBCO, we would require 1.5\% Zn or 1.0\% Ni.  Our measurements of the
magnetization in a magnetic field of 5 T, rule out the latter.

We propose that the broadening of the $^{17}$O(1) spectrum originates from local
moments in the copper oxygen plane that are associated with oxygen doping.  The exact placement of
the non-stoichiometric oxygen is still an open question\cite{McE05}; nonetheless, we conjecture
that this dopant modifies the spin state of an adjacent copper atom in the CuO$_2$ plane thereby
introducing a local moment in a manner similar to that which was proposed to account for local moment
formation in YBCO.\cite{All91,Bob97,Bob99} The
attendant inhomogeneous electronic structure is transferred through the hyperfine interaction to
the oxygen nucleus in the CuO$_2$ plane.  It was argued previously
in the YBCO work that an RKKY mechanism would produce positive and negative excursions of the local
magnetic field giving a symmetrically broadened NMR spectrum.  This feature is also a
characteristic aspect of our data.  McElroy {\it et al.}\cite{McE05} use STM
imaging spectroscopy to find disorder in the electronic structure of pure Bi-2212, 
specifically associated with the oxygen dopant and directly correlated to local  suppression 
of the superconducting coherence peaks.  Our interpretation of the $^{17}$O NMR spectra
is that the electronic disorder observed by STM  has an associated magnetic moment that forms in
the CuO$_2$ plane.  However, there are several issues that remain unresolved.  First, it is not
clear how such intrinsic moments form from non-stoichiometric oxygen.  Secondly,
the RKKY mechanism generally requires electron states far from the Fermi surface and should be
relatively insensitive to superconductivity which affects only those states on the gap scale,
close to the Fermi energy, E$_{F}$. Our observation of a strongly temperature dependent narrowing
of the $^{17}$O NMR spectra  in the superconducting state would be compatible with an RKKY
mechanism only if superconductivity modifies the density of states far from E$_{F}$.  In fact there is
some indication that this might be the case.  In their tunneling measurements McElroy {\it et al.}  find 
$dI/dV$,  immediately above the oxygen dopant atom, to be modified by superconductivity up to a bias of
$\approx$ -200 mV from E$_{F}$, which may be sufficient to suppress the RKKY interaction.

In conclusion, we have found a simple relation between the temperature dependent Knight
shift and NMR linewidth of $^{17}$O(1) spectra in Bi-2212 single crystals that indicates the existence of an
inhomogeneous magnetic field distribution from local moments in the the CuO$_2$ plane. We find a
Curie-law  behavior for the linewidth that decreases with increased oxygen doping.  Consequently we
identify the local moments with the oxygen dopant, intrinsic to Bi-2212, and we find they persist in the
superconducting state. However, the mechanism for formation of the magnetic moments is not known and
the effect of superconductivity on the hyperfine coupling between these moments and the probe nucleus,
$^{17}$O, needs additional clarification.

We are grateful for discussions with  A.V. Balatsky, H. Alloul, H. Cheng, J.C. Davis, P.J. Hirschfeld, A.
Trokiner, and A. Yazdani and contributions from  A.P. Reyes, P.L. Kuhns, V.F. Mitrovic, E.E Sigmund, and P.
Sengupta. This work was supported by the Department of Energy, grant DE-FG02-05ER46248.  P.G. and D.G.H.
acknowledge support for crystal growth from NSF-DMR-0449969 (PG), UWM-RGI-101X023 (PG), 
DOE-BES W-31-109-ENG-38, and NSF-DMR-91-20000.

\end{document}